\documentclass{ws-p8-50x6-00}
\begin{document}
\title{Theory of the Nucleon Spin-Polarizabilities II}
\author{Thomas R. Hemmert}
\address{IKP (Theorie), FZ J{\" u}lich, D-52425 J{\" u}lich, 
Germany\footnote{New address: 
Physik Department T39, TU M{\" u}nchen, D-85747
Garching, Germany.} \\
Email: th.hemmert@fz-juelich.de}

\maketitle

\abstracts{I discuss a generalization of the concept of nucleon 
spin-polarizabilities and present an overview of the available theoretical 
predictions up to now for these elusive but interesting structure 
parameters of the nucleon.}

The concept of nucleon spin-polarizabilities has been discussed in full detail
in the plenary talk contribution by Ulf Mei{\ss}ner \cite{Meissner}. In 
this brief contribution I want to point out that an interpretation of 
the 4 so called Ragusa spin-polarizabilities $\gamma_i$ \cite{Ragusa}
in terms of multipole
excitations and de-excitations can be given, if one resorts to a different
basis $R_i$ for the excitation part of the 6 structure functions $A_i$ of 
nucleon Compton scattering discussed in Mei{\ss}ner's talk.
For example \cite{BGLMN}, choosing the following linear 
combinations of the $\gamma_i$
\begin{eqnarray}
\gamma_{E1}&=&-\gamma_{1}-\gamma_{3}\quad
        \left(E1\rightarrow E1\right)\label{eq:1}\\
\gamma_{ M1}&=&\gamma_{4}\quad
        \left( M1\rightarrow M1\right)\\
\gamma_{ E2}&=&\gamma_{2}+\gamma_{4}\quad
        \left( M1\rightarrow E2\right)\\
\gamma_{ M2}&=&\gamma_{3}\quad
        \left(E1\rightarrow M2\right), \label{eq:4}
\end{eqnarray}
the resulting spin-polarizabilities $\gamma_{E1\,(M1)}$ can be interpreted
as the nucleon's spin response to an electric (magnetic) dipole excitation
via the incoming Compton photon and a successive dipole
de-excitation through the
outgoing Compton photon. Likewise, the other 2 linear combinations 
$\gamma_{E2\,(M2)}$ describe
the response to an electric (magnetic) dipole excitation and a successive
magnetic (electric) quadrupole de-excitation. The 4 Ragusa 
spin-polarizabilities can therefore be expressed in a basis where their
physical meaning becomes clear---these polarizabilities are just parameterizing
the leading spin response of the nucleon
for dipole excitations in (spin-dependent) Compton
scattering. As discussed by Mei{\ss}ner 
\cite{Meissner}, none of these (dipole) 
spin-polarizabilities has been measured directly up to know---results of 
future polarized Compton scattering experiments on the proton are eagerly 
awaited. In chiral perturbation theory the dipole spin-polarizabilities 
have now been calculated
to next-to-leading order, the results and a comparison with recent dispersion
theoretical calculations are given in table 1.

\begin{table}[t]
\caption{Predictions for the isoscalar and isovector dipole
spin-polarizabilities of the nucleon found to ${\cal O}(p^4)$ in heavy baryon
chiral perturbation theory (GHM), to ${\cal O}(\epsilon^3)$ in the small 
scale expansion (SSE1) and in dispersion analyses (DKH,DGPV,BGLM). All results
are given in the units $10^{-4}\,\rm{fm}^4$. 
($*$: $+\,2.5\times10^{-4}\,\rm{fm}^4$ from $\Delta$(1232) pole 
still missing)\label{tab:exp}}
\begin{center}
\footnotesize
\begin{tabular}{c||c|ccc|c}
$\gamma_{i}^{(N)}$ & GHM\cite{GHM} & DKH\cite{DKH} & 
DGPV\cite{DGPV} & BGLMN\cite{BGLMN} & SSE1\cite{SSE1}  \\
\hline
$\gamma_{E1}^{(s)}$ & $-3.0$ & $-5.0$ & $-5.2$ & $-4.5$ & $-5.2$\\
$\gamma_{M1}^{(s)}$ & $+0.4^{{*}}$ & $+3.4$ & $+3.4$ & $+3.3$ & $+1.4$\\
$\gamma_{E2}^{(s)}$ & $+2.0$ & $+2.4$ & $+2.7$ & $+2.4$ & $+1.0$\\
$\gamma_{M2}^{(s)}$ & $+0.6$ & $-0.6$ & $-0.5$ & $-0.2$ & $+1.0$\\
\hline
$\gamma_{E1}^{(v)}$ & $+1.2$ & $+0.5$ & $+0.8$ & $+1.1$ & - \\
$\gamma_{M1}^{(v)}$ & $+0.0$ & $+0.0$ & $-0.5$ & $-0.6$ & - \\
$\gamma_{E2}^{(v)}$ & $-0.2$ & $-0.2$ & $-0.5$ & $-0.5$ & - \\
$\gamma_{M2}^{(v)}$ & $+0.1$ & $-0.0$ & $+0.5$ & $+0.5$ & - \\
\hline
\end{tabular}
\end{center}
\end{table}

After realizing the possibility of such a multipole interpretation for
the Ragusa spin-polarizabilities, one of 
course does not have 
to stop classifying the nucleon's spin-dependent response in external 
electromagnetic fields at the level of
electric or magnetic dipole excitations. Recently, Holstein et al. \cite{HDPV}
have generalized such an analysis to classify spin structure responses
related to low energy quadrupole excitations of the nucleon. A straightforward
derivation of these higher order spin-polarizabilities can be given via
a multipole expansion of the 6 Compton structure amplitudes  
$R_i(\omega,\theta),\,i=1\ldots6$, where $\omega$ corresponds to the photon
energy in the center-of-mass system and $\theta$ denotes the cms 
scattering angle. 
For example, for {$i=6$} one finds 
\cite{BGLMN}:
\begin{eqnarray} R_6&=&\sum_{l\geq1}\left\{\left[
       f_{MM}^{l+}-f_{MM}^{l-}\right]\left(l\,P_{l}^{
       \prime\prime}+
       P_{l-1}^{\prime\prime\prime}\right)-\left[f_{EE}^{l+}-
       f_{EE}^{l-}\right]P^{\prime\prime\prime}_l\right.\nonumber \\
& &\phantom{\sum_{l\geq1}}
       -\left[f_{EE}^{l+}-
       f_{EE}^{l-}\right]P^{\prime\prime\prime}_l
       +f_{ME}^{l+}\left[\left(3l+1\right)P^{\prime\prime}_{l}+2\,
       P_{l-1}^{\prime\prime\prime}\right]\nonumber \\
& &\phantom{\sum_{l\geq1}}\left.-2\,f_{EM}^{l+}\left[
       \left(l+1\right)
       P_{l+1}^{\prime\prime}+2\,P_l^{\prime\prime\prime}\right]
       \right\} \nonumber
\end{eqnarray}
where the $P_i^a(z)$ denote the $a-$th derivative of Legendre-Polynomial $P_i$
and $f_{XY}^l(\omega)$ are the energy-dependent expansion parameters of
the multipole series. 
If one now maps the one-particle-irreducible or excitation part of $R_6$ onto 
the corresponding $A_6$ structure amplitude discussed by Mei{\ss}ner 
\cite{Meissner} and stops the multipole series at $l=1$, one obtains
\begin{eqnarray}
A_6^{1PI}&=&-c \, R_6^{1PI}\,=+\,c\,12\,f_{EM}^{1+}+(l\geq2)\nonumber\\
         &=&4\,\pi\,\gamma_3\,\omega^3+\ldots,
\end{eqnarray}
where $c$ corresponds to an amplitude normalization constant and $\gamma_3$
denotes one of the spin-polarizabilities discussed by Mei{\ss}ner. We can
therefore make the identification
\begin{eqnarray}
\rightarrow\gamma_3&=&\frac{3c\,f_{EM}^{1+}(\omega)}{\omega^3\pi}|_{\omega
\rightarrow 0}
\equiv\gamma_{M2}\quad\quad{\left(E1\rightarrow M2\right)}, \nonumber
\end{eqnarray}
which reproduces Eq.(\ref{eq:4}). Generalizing this analysis to $l=2$ one
obtains a set of 12 spin-polarizabilities defined via the expansion
parameters $f_{XY}^l(\omega)$ introduced above:
\begin{eqnarray}
4\,\pi\,\omega^3\left({\gamma_{E1}}+\omega^2
{\gamma_{E1,\nu}}\right)=&f_{EE}^{1+}(\omega)-f_{EE}^{1-}(\omega)
&(E1\rightarrow E1)\nonumber\\
4\,\pi\,\omega^3\left({\gamma_{M1}}+\omega^2
{\gamma_{M1,\nu}}\right)=&f_{MM}^{1+}(\omega)-f_{MM}^{1-}(\omega)
&(M1\rightarrow M1)\nonumber\\
4\,\pi\,\omega^3\left({\gamma_{E2}}+\omega^2
{\gamma_{E2,\nu}}\right)=&6\,f_{ME}^{1+}(\omega)
&(M1\rightarrow E2)\nonumber\\
4\,\pi\,\omega^3\left({\gamma_{M2}}+\omega^2
{\gamma_{M2,\nu}}\right)=&6\,f_{EM}^{1+}(\omega)
&(E1\rightarrow M2)\nonumber\\
4\,\pi\,\omega^5\,{\gamma_{ET}}\,=&3\left[f_{EE}^{2+}(\omega)-f_{EE}^{2-}(
\omega)\right]
&(E2\rightarrow E2)\nonumber\\
4\,\pi\,\omega^5\,{\gamma_{MT}}\,=&3\left[f_{MM}^{2+}(\omega)-f_{MM}^{2-}(
\omega)\right]
&(M2\rightarrow M2)\nonumber\\
4\,\pi\,\omega^5\,{\gamma_{E3}}\,=&15\,f_{ME}^{2+}(\omega)
&(M2\rightarrow E3)\nonumber\\
4\,\pi\,\omega^5\,{\gamma_{M3}}\,=&15\,f_{EM}^{2+}(\omega)
&(E2\rightarrow M3)\nonumber .
\end{eqnarray}
The 4 dipole spin-polarizabilities $\gamma_{E1},\gamma_{M1},\gamma_{E2},
\gamma_{M2}$ have already been discussed in Eqs.(\ref{eq:1}-\ref{eq:4}). When 
extending the multipole series to $l=2$, one in addition also picks up the
4 slopes $\gamma_{E1,\nu},\gamma_{M1,\nu},\gamma_{E2,\nu}, \gamma_{M2,\nu}$ of 
their energy-dependence, which are usually called dispersion 
spin-polarizabilities. Finally, there are 4 proper higher order 
spin-polarizabilities $\gamma_{ET},\gamma_{MT},\gamma_{E3},\gamma_{M3}$,
which correspond to quadrupole excitations and successive quadrupole oder
even octupole de-excitations.
Results of recent predictions for these essentially unknown higher order 
spin structure
parameters of the nucleon obtained in chiral effective field theories
and via dispersion analyses are given in table 2. It is interesting to note 
that all predictions agree that the dispersive effects of the dipole
spin-polarizabilities should give the dominant effect among the $l=2$
spin-polarizabilities---quadrupole and octupole de-excitation only seems
to play a minor role. This is another manifestation of the rather rigid
behavior of the nucleon under deformation forces as in the case of the
small quadrupole $N\Delta$-transition moment or the rather small electric 
(magnetic) polarizabilities $\bar{\alpha}_{E},\,(\bar{\beta}_M)$. 
Furthermore, the next-to-leading order chiral perturbation theory calculation 
gives the first indication that as in the case of the dipole 
spin-polarizabilities (table 1)
also for the higher order spin-polarizabilities the 
isoscalar components
clearly dominate over the isovector part (table 2), 
making proton and neutron rather
similar objects as far as their spin-structure at low energies is concerned. 

I would like to thank the organizer of GDH2000
for the opportunity to
present these new results 
and I am grateful for their financial support.

\begin{table}[t]
\caption{Status of the theoretical predictions as of summer 2000 for the 
higher order isoscalar and isovector spin-polarizabilities as given by
next-to-leading order chiral perturbation theory, by dispersion analyses
and to leading order in the small scale expansion. All results are given
in the units $10^{-4}\,\rm{fm}^6$. Note that the chiral next-to-leading 
calculations for the higher order isovector spin-polarizabilities find
$\gamma_{ET}^{(v)}=\gamma_{MT}^{(v)}=\gamma_{E3}^{(v)}=\gamma_{M3}^{(v)}=0$,
while no numbers from dispersion theory are available yet.
($*$: $+\,1.20\times10^{-4}\,\rm{fm}^6$ from $\Delta$(1232) pole
still missing)\label{tab2}}
\begin{center}
\footnotesize
\begin{tabular}{c||c|c|c|}
$\gamma_{i}^{(N)}$ & ${\cal O}(p^4)$ ChPT\cite{GHHM} & Disp. Th.\cite{HDPV} & 
${\cal O}(\epsilon^3)$ SSE\cite{HDPV} \\
\hline
$\gamma_{ET}^{(s)}$ & $+0.08$ & $-0.15$ & $-0.28$\\
$\gamma_{MT}^{(s)}$ & $+0.06$ & $-0.09$ & $-0.03$\\
$\gamma_{E3}^{(s)}$ & $+0.03$ & $+0.06$ & $+0.11$\\
$\gamma_{M3}^{(s)}$ & $+0.03$ & $+0.09$ & $+0.11$\\
\hline
$\gamma_{E1,\nu}^{(s)}$ & $-3.40$ & $-3.42$ & $-5.16$\\
$\gamma_{M1,\nu}^{(s)}$ & $+0.63^{*}$ & $+2.23$ & $+0.83$\\
$\gamma_{E2,\nu}^{(s)}$ & $+1.68$ & $+1.30$ & $+0.28$\\
$\gamma_{M2,\nu}^{(s)}$ & $-0.12$ & $-0.60$ & $-0.22$\\
\hline\hline
$\gamma_{E1,\nu}^{(v)}$ & $+0.97$ & ? & -\\
$\gamma_{M1,\nu}^{(v)}$ & $+0.00$ & ? & -\\
$\gamma_{E2,\nu}^{(v)}$ & $-0.09$ & ? & -\\
$\gamma_{M2,\nu}^{(v)}$ & $+0.05$ & ? & -\\
\hline
\end{tabular}
\end{center}
\end{table}

\end{document}